\documentclass[elsarticle.cls,reprint,showpacs,showkeys,superscriptaddress,preprintnumbers,groupedaddress]{revtex4-1}
\usepackage{amsmath,amssymb,bm,mathrsfs}
\usepackage{srcltx}
\usepackage{dsfont}
\usepackage{epsfig}
\usepackage{slashed}
\usepackage{bbold}
\usepackage{psfrag}
\usepackage{xcolor}

\usepackage{subfig}
\usepackage{xfrac}
\usepackage{multirow}
\usepackage{booktabs}
\usepackage[colorlinks=true,linkcolor=blue,citecolor=blue,urlcolor=violet]{hyperref}
\usepackage{youngtab}

\newcommand{\wb}{\overline}

\usepackage{soul}

\begin{document}

\title{
The Octagon and the Non-Supersymmetric String Landscape
}

\author{Riccardo Argurio$^1$}
\email{rargurio@ulb.ac.be}
\author{Matteo Bertolini$^{2,3}$}
\email{bertmat@sissa.it}
\author{Sebasti\'an Franco$^4$}
\email{sfranco@ccny.cuny.edu}
\author{Eduardo Garc\'ia-Valdecasas$^1$}
\email{eduardo.garcia.valdecasas@gmail.com}
\author{Shani Meynet$^2$}
\email{smeynet@sissa.it}
\author{Antoine Pasternak$^1$}
\email{antoine.pasternak@ulb.ac.be}
\author{Valdo Tatitscheff$^5$}
\email{valdotatitscheff@gmail.com}

\affiliation{$^1$ Physique Th\'eorique et Math\'ematique and International Solvay Institutes, \\ Universit\'e Libre de Bruxelles; C.P. 231, 1050 Brussels, Belgium
} 
\affiliation{$^2$ SISSA and INFN - Sezione di Trieste \\
Via Bonomea 265; I 34136 Trieste, Italy
} 
\affiliation{$^3$ ICTP - Strada Costiera 11; I 34014 Trieste, Italy
}
\affiliation{$^4$ Physics Department, The City College of the CUNY \\160 Convent Avenue, New York, NY 10031, USA
} 
\affiliation{$^5$
IRMA, UMR 7501, Universit\'e de Strasbourg et CNRS \\ 7 rue Ren\'e Descartes 67000 Strasbourg, France
}

\begin{abstract}
We present an orientifold of a toric singularity allowing for a configuration of fractional branes which corresponds to a gauge theory that dynamically breaks supersymmetry in a stable vacuum. 
This model represents the first such instance within the gauge/gravity duality.

\end{abstract}

\maketitle

\noindent
{\em Introduction---}%
Our understanding of gauge theories was given a dramatically new perspective when it was realized that they appear ubiquitously in string theory. In particular, four-dimensional gauge theories with ${\cal N}=1$ supersymmetry can be engineered by considering D-branes at Calabi-Yau three-fold singularities in type IIB string theory, in the limit in which supergravity decouples from the open string degrees of freedom \cite{Klebanov:1998hh}. For those singularities which are toric, there is a specific algorithm that completely determines the gauge theory, modulo Seiberg dualities. The gauge theories one gets are usually of quiver type, with all their data (gauge groups, chiral matter fields and superpotential couplings) best encoded in a dimer model \cite{Franco:2005rj,Franco:2005sm}.

It is a question of interest to ask whether all kinds of ${\cal N}=1$ supersymmetric gauge theories can be engineered in this way, or at least if it is possible to engineer theories which reproduce all kinds of low-energy behavior. While confinement, generation of a mass gap and of a chiral condensate can be shown to arise in very simple models \cite{Klebanov:2000hb}, as well as ${\cal N}=2$ Coulomb-like branches in others \cite{Bertolini:2000dk,Polchinski:2000mx,Bertolini:2001gg}, the fascinating possibility that the vacuum of the gauge theory dynamically breaks supersymmetry requires more work. 

Supersymmetry can be broken in different ways. The gauge theory may have both supersymmetric vacua and meta-stable supersymmetry breaking vacua, which can be parametrically long-lived. This situation can be engineered with D-branes at singularities, see {\it e.g.} \cite{Kachru:2002gs,Franco:2006es,Argurio:2006ny,Argurio:2007qk}. Another possibility is that there is simply no vacuum in the theory, leading to what is called a runaway. It turns out that such a situation is rather frequent in configurations of branes at singularities, see \cite{Berenstein:2005xa,Franco:2005zu,Bertolini:2005di,Intriligator:2005aw}.

The last possibility that remains is that supersymmetry is dynamically broken in a fully stable vacuum. This has proven to be a harder problem to engineer with D-branes at singularities. This is partly due to the scarcity of known gauge theories that display such a non-supersymmetric vacuum. After attempts to turn the runaway into a stable vacuum proved unsuccessful \cite{Argurio:2006ew}, 
it was shown in \cite{Franco:2007ii} that  introducing an orientifold projection it is possible to engineer configurations which at low-energies reproduce the well-known dynamical supersymmetry breaking (DSB) model of \cite{Affleck:1983vc}, henceforth referred to as `the $SU(5)$ model.' The same model was argued to arise in a wider number of singularities in \cite{Retolaza:2015nvh,Retolaza:2016alb}. Besides its intrinsic interest, the existence of such models is also important because it could be in tension with recent conjectures such as the one presented in \cite{Buratti:2018onj} and, more generally, can be of some relevance within the  swampland program \cite{Vafa:2005ui,Brennan:2017rbf,Palti:2019pca}.

Somewhat in a plot twist, the DSB configurations of \cite{Franco:2007ii,Retolaza:2015nvh} were more closely scrutinized in \cite{Buratti:2018onj}, where it was found that they are actually not fully stable. Indeed, when the DSB configuration is probed by $N$ regular D3-branes, an instability appears where the regular branes split along the Coulomb branch of so-called ${\cal N}=2$ fractional branes \cite{Franco:2005zu},  eventually settling the configuration in a supersymmetric vacuum. This phenomenon was further investigated in \cite{Argurio:2019eqb}, where many examples of brane configurations at orientifolded singularities with a DSB model were found, all with the same kind of instability. In fact, a no-go theorem was proven in \cite{Argurio:2019eqb} showing that for any singularity allowing for a DSB model, ${\cal N}=2$ fractional branes, if present, always destabilize the supersymmetry breaking vacuum and set, eventually, the vacuum energy to zero. All of this was mounting evidence for what could be interpreted as the impossibility of engineering stable DSB with D-branes at singularities. In more dramatic words, could stable DSB be in the swampland?

In this letter we argue that stable DSB is still in the landscape. We produce an orientifold of a toric singularity which allows for a brane configuration displaying a variant of the $SU(5)$ DSB model, and that has no instabilities. In particular,  those  described in \cite{Buratti:2018onj,Argurio:2019eqb} are absent because the singularity does not admit ${\cal N}=2$ fractional branes. This provides a counter-example to what could have been conjectured, namely that DSB models were possible only in singularities admitting ${\cal N}=2$ fractional branes, and hence, following the no-go theorem presented in \cite{Argurio:2019eqb}, unstable towards supersymmetric vacua \footnote{We do not investigate here the existence of stable DSB models in brane constructions that include flavors, engineered by non-compact D7-branes, for which it is already known that metastable DSB vacua can be found \cite{Franco:2006es}.}.

\begin{figure}[!!!t]
	\begin{center}
		\includegraphics[width=0.20 \textwidth]{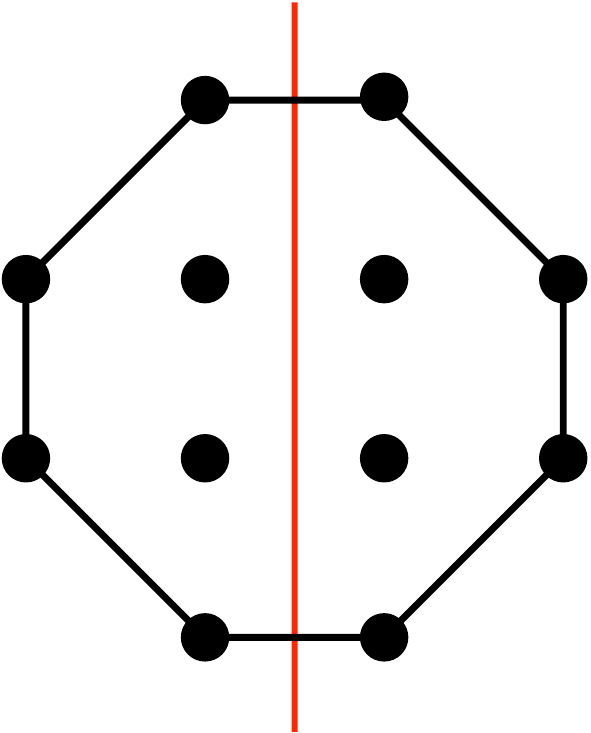}
	\end{center}
	\vspace{-.3cm}
	\caption{\small The toric diagram of the Octagon singularity.
		\label{fig:toricdiagram}}
\end{figure}

\vspace{.2cm}
\noindent
{\em The Octagon---}%
The toric singularity we start with is the following. We dub it the `Octagon' because of its toric diagram, that we reproduce in Fig.~\ref{fig:toricdiagram}. It has 8 edges and it is of area 14, where the unit of area is an elementary triangle with which one performs a triangulation of the diagram. Before any orientifold projection, D-branes probing this singularity lead to a gauge theory with 14 gauge groups. 

The four-dimensional ${\cal N}=1$ gauge theories living on the worldvolume of (fractional) D3-branes probing toric Calabi-Yau three-fold singularities are fully encoded by bipartite graphs on a two-torus known as dimer models or brane tilings \cite{Franco:2005rj,Franco:2005sm}.  A simple dictionary connects dimers to the corresponding gauge theories. Faces, edges and nodes in the dimer correspond to gauge group factors, bi-fundamental or adjoint chiral fields and superpotential terms, respectively. Dimers significantly simplify the connection between the geometry of the singularity and the corresponding gauge theory. Moreover, dimers efficiently encode orientifolds, which translate into $\mathbb{Z}_2$ involutions of the graph. We will focus on the class of involutions studied in \cite{Franco:2007ii}, which have either fixed points or fixed lines \footnote{In principle, $\mathbb{Z}_2$ actions without fixed loci are possible, but they have not been investigated in the literature.}.

\begin{figure}[!!!t]
\begin{center}
\includegraphics[width=0.34 \textwidth]{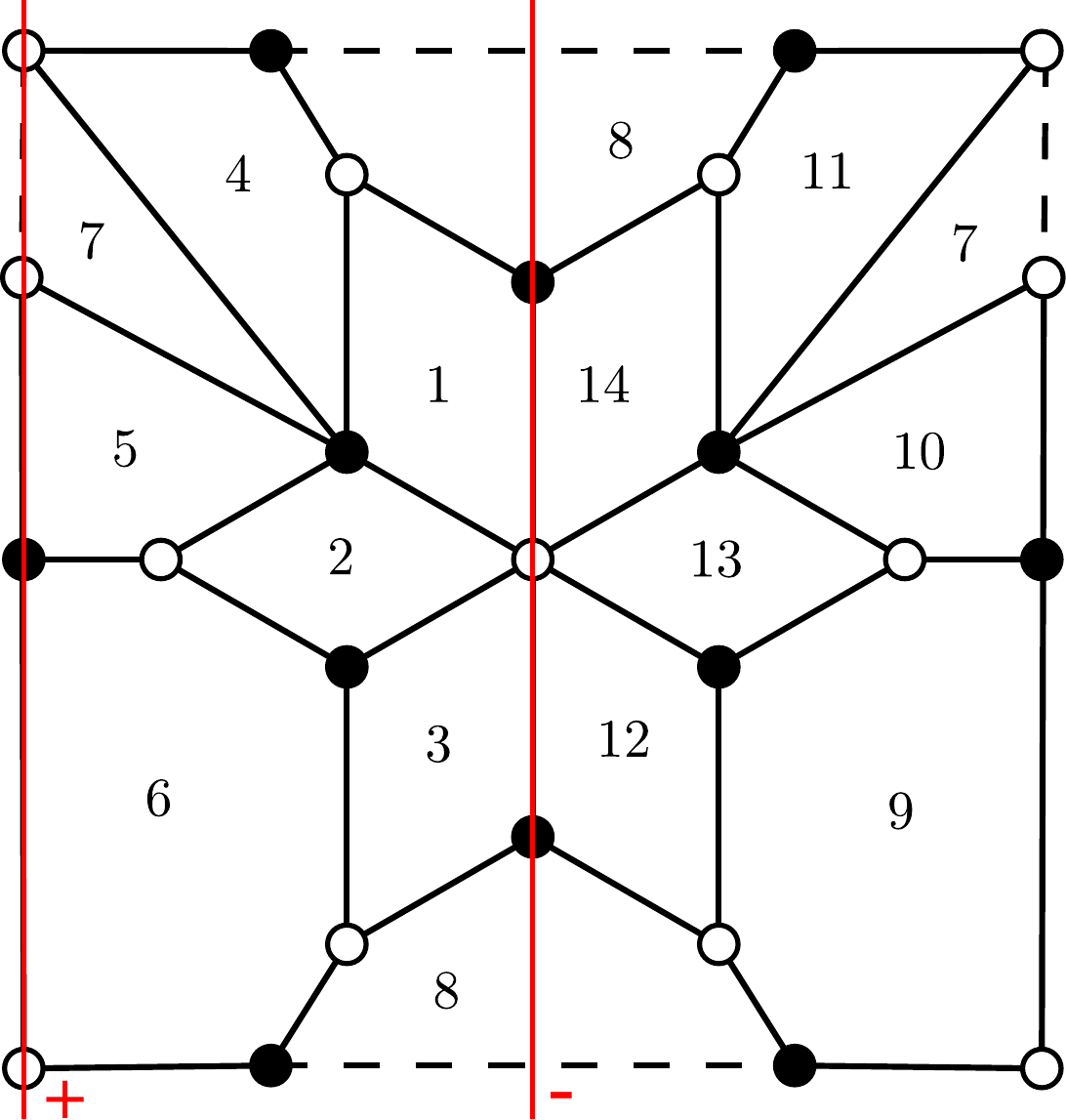}
\end{center}
\vspace{-.3cm}
\caption{\small The unit cell of the dimer of the Octagon with orientifold lines (in red).
\label{fig:dimer}}
\end{figure}

According to the rules stated in \cite{Retolaza:2016alb}, the Octagon does not admit any orientifold represented as a point projection on the dimer. On the other hand, the highly symmetric toric diagram shows that orientifolds represented by line projections are possible, both diagonal and vertical/horizontal (the latter two possibilities are obviously equivalent). In Fig.~\ref{fig:dimer} we show the dimer of the Octagon, together with its two orientifold vertical lines. More precisely, this is the dimer corresponding to a particular toric phase where the vertical fixed lines are manifest. Other toric phases, obtained by Seiberg dualities, obviously exist, but in general do not display the symmetry required to perform the (vertical) line projection. That the dimer of Fig.~\ref{fig:dimer} indeed corresponds to the toric singularity of Fig.~\ref{fig:toricdiagram} can be checked using standard techniques \cite{Franco:2005rj,Franco:2005sm}.

In an orientifold with vertical fixed lines, each line carries an independent sign, which controls the projections of gauge groups and matter fields.
The orientifold lines identify the 6 faces (1--6) to the 6 faces (14--9), respectively, each corresponding to an $SU$ gauge group. Faces 7 and 8 are self-identified. By assigning the sign $+$ to the line on the edge of the unit cell, and the sign $-$ to the line in the middle of the cell, faces 7 and 8 inherit an $SO$ and an $USp$ gauge group, respectively. Moreover, the $SU$ groups of faces 1 and 3 have a matter field in the antisymmetric representation, while the $SU$ groups of faces 5 and 6 have a matter field in the symmetric representation.

Before performing the orientifold projection, it is straightforward to see that the following rank assignment is anomaly free: faces $1, 2, 3, 7, 12, 13$ and $14$ have gauge group $SU(N+M)$, and all the others have gauge group $SU(N)$. Setting $N=0$, one has only seven $SU(M)$ gauge groups: one isolated Super-Yang-Mills (SYM) on face 7 and the six others forming a loop whose links are bi-fundamentals, together with a sextic superpotential proportional to the only gauge invariant (it is represented by the white dot in the center of the unit cell).  This rank assignment corresponds to a so-called deformation fractional brane \cite{Franco:2005zu}. One can easily see that such a gauge theory eventually leads to a confining behavior just like SYM. This can be naturally UV completed starting from a system of $N$ regular and $M$ fractional D3-branes which trigger a RG-flow that can be described by a duality cascade, similar to  \cite{Klebanov:2000hb} and many other examples that were found since then. The effective number of regular branes diminishes along the flow and the deep IR dynamics is described by fractional branes only.

In the presence of an orientifold projection, it is no longer granted that an anomaly free rank assignment exists at all. For instance, in the present case it can be shown that it is not possible to find one if the signs of the two lines are the same. However, choosing opposite signs as in Fig.~\ref{fig:dimer}, one can see that there is a rank assignment which is anomaly free: $SU(N+M+4)$ for faces 1 and 3, $SU(N+M)$ for face 2, $SO(N+M+4)$ for face 7, $SU(N)$ for faces 4, 5 and 6, and $USp(N)$ for face 8. Setting $N=0$ we obtain a gauge theory with an isolated $SO(M+4)_7$ SYM theory, which confines on its own, together with a quiver gauge theory based on the group $SU(M+4)_1\times SU(M)_2 \times SU(M+4)_3$ with matter fields and a superpotential that we proceed to analyze. 

\vspace{.2cm}
\noindent
{\em The DSB model---}%
The gauge theory 
\begin{equation}
SU(M+4)_1\times SU(M)_2 \times SU(M+4)_3 
\end{equation}
has matter content
\begin{equation}
A_1 = {\tiny {\yng(1,1)}_{\,1}}, \
X_{12}=(\wb{\tiny  \yng(1)}_{\, 1},{\tiny  \yng(1)}_{\, 2}) , \ 
X_{23}=(\wb {\tiny  \yng(1)}_{\, 2},  {\tiny  \yng(1)}_{\, 3} ),\  A_3 = {\tiny \wb {\yng(1,1)}_{\,3}}
\end{equation}
and superpotential
\begin{equation}
W=A_1 X_{12}X_{23}A_3 X_{23}^t X_{12}^t\ .
\end{equation}
The superpotential can be interpreted as follows. The gauge invariant $ X_{12}^tA_1 X_{12}$ of group 1 and the gauge invariant $X_{23}A_3 X_{23}^t$ of group 3 are respectively in the ${\tiny {\yng(1,1)}_{\,2}}$ and ${\tiny \wb {\yng(1,1)}_{\,2}}$ of gauge group 2, with $W$ above providing a bilinear in these two invariants, thus akin to a mass term. It is obvious that the antisymmetrics of $SU(M)_2$ can exist as such only if $M\geq 2$. In this case, one can show that strongly coupled dynamics generates superpotential terms that, together with the tree level one, eventually lead to supersymmetric vacua. For $M = 0$ one gets instead two decoupled theories at faces 1 and 3 both having gauge group $SU(4)$ and one chiral superfield in the antisymmetric, which have a runaway behavior. The case of interest is $M=1$.

For $M=1$ node 2 becomes trivial ($SU(1)$ is empty) and, more importantly, the superpotential actually vanishes. Indeed, both nodes 1 and 3 are $SU(5)$ gauge theories with matter in the ${\tiny {\yng(1,1)}}\oplus\wb{\tiny  \yng(1)} $ representations, and there is no chiral gauge invariant that can be written in this situation \cite{Affleck:1983vc}. Hence the two gauge theories are effectively decoupled, and their IR behavior can be established independently. Both happen to be the $SU(5)$ model for stable DSB. Since the $SO(5)$ SYM on node 7 just confines, we thus determine that this configuration displays DSB in its vacuum. Quite interestingly, this DSB vacuum may then arise at the bottom of a duality cascade (possibly more complicated with respect to the simpler unorientifolded case, due to the orientifold projection which would modify it, see \cite{Argurio:2017upa}), hence within a stringy UV completed theory.

\vspace{.2cm}
\noindent
{\em Stability---}%
Is this DSB vacuum stable? In principle, there can be different sources of potential instabilities. 

First, one could be concerned about stringy instantons, whose presence may affect the low energy dynamics. Indeed, the D-brane configuration giving rise to the twin $SU(5)$ DSB model, $N=0,M=1$, contains both a $USp(0)$ and an $SU(1)$ factor coupling to the $SU(5)$ gauge groups.  These are the two instances where contributions to the low-energy effective superpotential are allowed (see \cite{Argurio:2007vqa} and \cite{Petersson:2007sc}, respectively). However, no such contributions can be generated in our model simply because there are no chiral gauge invariants that can be written which can contribute to the superpotential. We thus conclude that stringy instantons cannot alter the DSB dynamics.

A second source of instability is the one discussed in \cite{Buratti:2018onj,Argurio:2019eqb}. In fact, as can be readily seen from the toric diagram of Fig.~\ref{fig:toricdiagram}, this singularity does not admit ${\cal N}=2$ fractional branes. The latter arise when the singularity can be partially resolved to display, locally, a non-isolated $\mathbb{C}^2/\mathbb{Z}_n$ singularity and a Coulomb-like branch associated to it. This translates into the presence of points inside some of the edges along the boundary of the toric diagram. The Octagon does not have this property. Hence, without the presence of ${\cal N}=2$ fractional branes, there is no vacuum expectation value on which the energy of the DSB vacuum can depend on, or equivalently there is no Coulomb branch along which the energy can slide to zero value.  

A final source of instability may come from the ${\cal N}=4$ Coulomb branch represented by regular D3-branes. As in the previously analyzed cases \cite{Buratti:2018onj,Argurio:2019eqb}, one can easily show that this is a non-supersymmetric flat direction, essentially because of the conformality of the parent (non-orientifolded, large $N$) gauge theory. Therefore, there are no supersymmetric vacua along this branch \footnote{Flat directions are usually not expected in a non-supersymmetric vacuum. Subleading $1/N$ corrections to anomalous dimensions of matter fields, which could lift such flat direction, are not easily calculable, particularly in a complicated singularity such as the Octagon. However, they should neither change the number of supersymmetric vacua nor modify the behavior of the potential at infinity, at least for sufficiently large $N$.}. 

\vspace{.2cm}
\noindent
{\em Conclusions---}%
In this letter we have presented a model, based on the Octagon, which is the first instance, to our knowledge, of a stable DSB configuration of fractional branes. As an existence proof of such configurations, this is enough. However, it is not by chance that this particular singularity has been found, rather one can be led to it by a series of arguments. 
This is reviewed in \cite{Argurio:2020npm}, where it is also shown that the Octagon is in fact the simplest singularity allowing for stable DSB. Other, more complicated singularities may realize it, but always through the twin $SU(5)$ model, and this is the reason why the simplest occurrence of this phenomenon is a singularity corresponding to a quiver with no less than 14 gauge groups. More details on how to find such toric singularities, and subtleties regarding orientifold projections and anomaly cancellation conditions, appear in \cite{Argurio:2020dko}.

With this example, we have shown that stable DSB can still be engineered by brane configurations at Calabi-Yau singularities. Given the remarkable properties of this family of models, we consider it important to study them in further detail.

\newpage
\noindent
{\em Acknowledgements---} 
We are grateful to I\~naki Garc\'ia-Etxebarr\'ia and Angel Uranga for discussions and for enlightening comments on a preliminary draft version.  R.A., A.P. and E.G.-V. acknowledge support by IISN-Belgium (convention 4.4503.15) and by the F.R.S.-FNRS under the ``Excellence of Science" EOS be.h project n.~30820817, M.B. and S.M. by the MIUR PRIN Contract 2015 MP2CX4 ``Non-perturbative Aspects Of Gauge Theories And Strings" and by INFN Iniziativa Specifica ST\&FI. E.G.-V. was also partially supported by the ERC Advanced Grant ``High-Spin-Grav". The research of S.F. was supported by the U.S. National Science Foundation grants PHY-1820721 and DMS-1854179. R.A. is a Research Director and A.P. is a FRIA grantee of the F.R.S.-FNRS (Belgium).

\bibliographystyle{apsrev4-1}
\bibliography{biblio}

\end{document}